\documentclass[prl,aps,showpacs,superscriptaddress,amsmath,amssymb,floatfix,twocolumn]{revtex4}
\usepackage{graphicx}
\usepackage{color}

\begin{document}

\title{Double symmetry breaking and 2D quantum phase diagram in spin-boson systems}
\date{\today} 

\author{Pierre Nataf}
\email{pierre.nataf@univ-paris-diderot.fr}
\author{Alexandre Baksic}
\author{Cristiano Ciuti}
\email{cristiano.ciuti@univ-paris-diderot.fr}
\affiliation{Laboratoire Mat\'eriaux et Ph\'enom\`enes Quantiques,
Universit\'e Paris Diderot-Paris 7 et CNRS, \\ B\^atiment Condorcet, 10 rue
Alice Domon et L\'eonie Duquet, 75205 Paris Cedex 13, France}

\begin{abstract}
The quantum ground state properties of two independent chains of spins (two-levels systems) interacting with the same bosonic field are theoretically investigated. Each chain is coupled to a different quadrature of the field, leading to two independent symmetry breakings for increasing values of the two spin-boson interaction constants $\Omega_C$ and $\Omega_I$. A  phase diagram is provided in the plane ($\Omega_C$,$\Omega_I$) with 4 different phases that can be characterized by the complex bosonic coherence of the ground states and can be manipulated via non-abelian Berry effects. In particular, when $\Omega_C$ and $\Omega_I$ are both larger than two critical values, the fundamental subspace has a four-fold degeneracy.  Possible implementations in superconducting or atomic systems are discussed. \end{abstract}

\maketitle

Spontaneous symmetry breaking is an important feature of quantum phase transitions\cite{sachdev} and typically implies the appearance of degenerate ground states. An understanding and control of the symmetries of a quantum system can lead  to the realization of new quantum phases for fundamental studies and/or for quantum applications.
One example is given by the Dicke model\cite{Dicke} describing the coupling of a collection of independent two-level systems coupled to a single-mode bosonic field. In the Dicke model, for a critical spin-boson coupling, a quantum phase transition occurs with the appearance of two-degenerate ground states\cite{emary}. Recently, such an Hamiltonian has been experimentally implemented in an effective way in a system where a pump-dressed Bose-Einstein condensate is embedded in an optical cavity\cite{ZurichNature,ZurichPRL}.
In the case of the Dicke model, the Hamiltonian symmetry is the parity of the total number of excitation quanta in the system  \cite{emary,Lehur}. The symmetry breaking across the transition means that the degenerate ground states (the so-called 'superradiant' phase) are not eigenstates of the parity operator $\Pi$, contrary to the non-degenerate ground state (normal phase) obtained for sub-critical coupling.  More generally, the spin-boson systems are currently attracting a great deal of attention, since they can be implemented with very controllable artificial systems such as dressed ultracold atoms in optical cavities and superconducting  circuit QED\cite{Wallraff}.

In this Letter, we explore the properties of a spin-boson Hamiltonian describing two independents chains of two-levels systems interacting with the same bosonic field in a way related, but qualitatively different to the Dicke model. One chain is coupled to a quadrature of a bosonic field, while the other chain is coupled to the orthogonal quadrature of the same field via two independent spin-boson coupling constants, which we will call  $\Omega_C$ and $\Omega_I$ in the following.  Like in the usual Dicke Hamiltonian,  the parity operator $\Pi$ commutes with the system Hamiltonian and describes a symmetry of the system. However, we will show that  $\Pi$ can be written as a product of $2$ symmetry operators which can be independently and spontaneously broken  if $\Omega_C$ and $\Omega_I$ are increased above the corresponding critical values $\Omega_C^{cr}$ and $\Omega_I^{cr}$.  We calculate the phase diagram and energy spectrum of the elementary bosonic excitations of the system in the thermodynamical limit.  A phase diagram with $4$ different zones in the $(\Omega_C,\Omega_I)$ plane  is provided and characterized in terms of the macroscopic coherences.
When both $\Omega_C>\Omega_C^{cr}$ and $\Omega_I>\Omega_I^{cr}$, the ground state subspace is four-fold degenerate, with complex bosonic coherences. 
We show that adiabatic loops in the  ($\Omega_C$,$\Omega_I$)  space provide non-abelian Berry geometric unitary transformations, which can be used to manipulate the four-fold degenerate ground states. Possible realizations in circuit QED and atomic systems are discussed.

The `double-chain' Dicke Hamiltonian under consideration is described by the following spin-boson Hamiltonian:
\begin{eqnarray}
\label{hamiltonian}
H/\hbar = \,\omega_{cav} a^{\dag}  a \,+ \,\omega^0_C J_z^C\,\,+ \,\omega^0_I J_z^I \,\,\,\,\\\nonumber + \frac{2\Omega_C}{\sqrt{N_C}}(a + a^{\dag})J_x^C\,+ i\frac{2\Omega_I}{\sqrt{N_I}}(a - a^{\dag}) J_x^I , \nonumber
\end{eqnarray} 
where $a^{\dagger}$  is the bosonic creation  operator with energy $\hbar \omega_{cav}$.
The total angular momentum operators $J_z^C$ and $J_x^C$ ($J_z^I$ and $J_x^I$) describe the ensemble of
$N_C$ ($N_I$)  two-levels systems with transition frequency $\omega^0_C$ ($\omega^0_I$). The first chain is coupled to the boson field quadrature $(a\,+\,a^{\dag})$, while the second is coupled to the quadrature $i(a\,-\,a^{\dag})$, via the coupling constants $\Omega_C$ and $\Omega_I$ respectively. The collective atomic operators are defined for $k\in\{I,C\}$ as  $J_z^k=\sum_{l_k=1}^{N_k} \sigma_z^{l_k} $ and $J_{x}^k=(1/2)(J_{+}^k+J_{-}^k)=(1/2)\sum_{l_k=1}^{N_k} (\sigma_{+}^{l_k}\,+\,\sigma_{-}^{l_k})$,
 with  $\sigma_z^{l_k}$ and $\sigma_{\pm}^{l_k}$ the usual Pauli matrices for the ${l_k}^{th}$ pseudo-spin, so that 
 the angular commutation  relations read $[J_z^k,J_{\pm}^{k'}]=\pm\delta_{k,k'}J_{\pm}^k\,\,;\,\,\,\,\,[J_+^k,J_-^{k'}]=2 \delta_{k,k'}J_{z}^k$  for $k,k'\in\{I,C\}$.

Let us start our analysis of such a `double chain' Dicke model by describing the role of the parity operator $\Pi_s$ in the standard `single chain' Dicke Hamiltonian $H_s/\hbar=\omega_{cav} a^{\dag}  a \,+ \,\omega^0J_z\,\,+\,\,\Omega/\sqrt{N}(a + a^{\dag})(J_+ + J_-)$ . The anti-resonant terms   $J_- a $ and  $a^{\dag}J_+$ prevent the conservation of the total number of excitations $J_z+N/2+a^{\dag}a$ (contrary to the Jaynes-Cumming or Tavis-Cummings models \cite{tavis}), however the parity of the excitation number operator $\Pi_s=\exp(i\pi(J_z+N/2+a^{\dag}a))$ still commutes with the Hamiltonian\cite{emary}. In particular, $\Pi_s$ transforms the operators in $H_s$ as follows:
 \begin{eqnarray}
 \Pi_s:(a, J_x)\longrightarrow \Pi_s(a, J_x) \Pi_s^{\dag}=(-a, -J_x)
  \end{eqnarray}
while $J_z$ and $a^{\dag} a$ remain unchanged.
Analogously, in the present `double chain' Dicke model , the parity of the total number of excitations $\Pi$ is conserved. The corresponding symmetry operator is defined as:
\begin{eqnarray}
 \Pi=\exp(i\pi( a^{\dag} a \,+ J_z^C+N_C/2+ J_z^I+N_I/2)).
 \end{eqnarray}
It is apparent that $\left[H, \Pi\right]=0$ because:
 \begin{eqnarray}
\Pi : (a, J_x^C, J_x^I)\longrightarrow \Pi(a, J_x^C, J_x^I) \Pi^{\dag}=(-a, -J_x^C, -J_x^I).\,\,\,\,\,\,\,\,
   \end{eqnarray}
 
The novelty given by the coupling to the two different quadratures is the possibility to express the parity operator as a product  $\Pi=\mathcal{T}_I\circ\mathcal{T}_C$ where $\mathcal{T}_I$ and  $\mathcal{T}_C$ are independently conserved and are defined through the following transformations:
\begin{eqnarray}
(a+a^{\dag},i(a-a^{\dag}),J_x^C,J_x^I)\stackrel{\mathcal{T}_I}{\rightarrow} (a+a^{\dag},-i(a-a^{\dag}),J_x^C,-J_x^I), \nonumber \,\,\,\,\,\\
(a+a^{\dag},i(a-a^{\dag}),J_x^C,J_x^I)\stackrel{\mathcal{T}_C}{\rightarrow} (-a-a^{\dag},i(a-a^{\dag}),-J_x^C, J_x^I)  \,\,\,\,\,\,\,\, \nonumber
\end{eqnarray}
where $ a^{\dag} a$, $J_z^I$ and $J_z^C$ remain unchanged by those two transformations.
$\mathcal{T}_I$ 	and $\mathcal{T}_C$ are one the dual of each other: each one flips the sign of the corresponding quadrature of the field, while keeping unchanged the other one.
 There is a simple geometric interpretation of those symmetries in the complex quadrature plane $(Re(\langle a\rangle), Im(\langle a\rangle))$. $\Pi$ is represented by a $\pi$-rotation with  respect to the origin in such plane,  $\mathcal{T}_C$ is a mirror reflection with respect to the imaginary axis, while $\mathcal{T}_I$ is a mirror reflection with respect to the real axis. 
As we will show, the symmetries associated to $\mathcal{T}_C$ and  $\mathcal{T}_I$ break down when the corresponding coupling constants $\Omega_C$ and $\Omega_I$, are increased across some quantum critical lines. 

\begin{figure}
\begin{center}
\includegraphics[width=250pt]{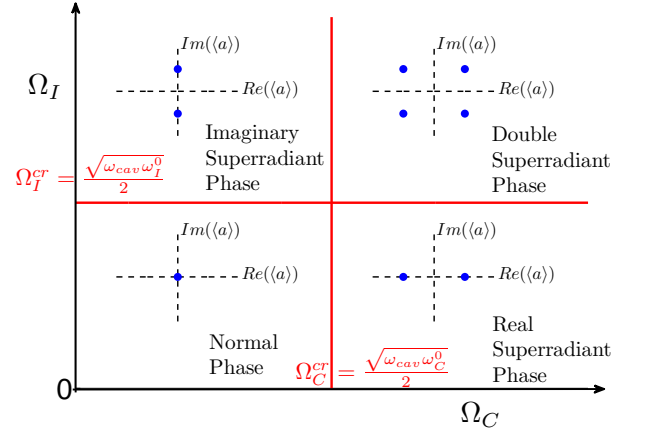}
\caption{Phase diagram of the `double chain' Dicke Hamiltonian in Eq. (\ref{hamiltonian}), exhibiting four different phases in the two-dimensional parameter $(\Omega_C,\Omega_I)$ space, where $\Omega_C$ ($\Omega_I$) is the spin-boson coupling for the first (second) chain of pseudospins (two-level systems).
For each of the $4$ phases, a sketch of the complex boson coherences $\langle a \rangle=\pm\sqrt{\gamma_C}+\mp i\sqrt{\gamma_I}$ for the ground states is depicted in the complex plane (blue points). In particular, for $\Omega_C>\Omega_C^{cr}$ and $\Omega_I>\Omega_I^{cr}$, the phase is characterized by 4 degenerate vacua, with $4$ different complex bosonic field coherences. Such coherences are mutually related via the symmetry operations $\mathcal{T}_C$ and $\mathcal{T}_I$ defined in the text.   \label{phasediagram}}
\end{center}
\end{figure}

To explore the properties of such an Hamiltonian, it is convenient to use the Holstein-Primakoff transformation \cite{holstein} which allows us to represent the angular momentum operators in terms of 2 independent bosonic modes: $J^k_+ = b^{\dag}_k (N_k - b^{\dag}_k  b_k)^{1/2}$ , $J^k_- = (N_k - b^{\dag}_k b_k)^{1/2} b_k$ and $J^k_z = b^{\dag}_k b_k\,-\,N_k/2$ for $k \in\{I,C\}$.
In this representation, the Hamiltonian reads:
\begin{eqnarray}
\label{Hbosonic}
H/\hbar = \,\omega_{cav} a^{\dag}  a \,+\,\omega^0_C(b^{\dag}_C b_C\,-\,\frac{N_C}{2})\,\,+ \,\omega^0_I ( b^{\dag}_I b_I\,-\,\frac{N_I}{2})\,\,\,\,\,\,\nonumber\\
 + \Omega_C(a + a^{\dag})\{ b^{\dag}_C \sqrt{1 - b^{\dag}_C b_C/N_C} + \sqrt{1- b^{\dag}_C b_C/N_C}\,\, b_C \}\,\nonumber\\+ i\Omega_I(a -a^{\dag})\{b^{\dag}_I \sqrt{1 - b^{\dag}_I b_I/N_I} + \sqrt{1-b^{\dag}_I b_I/N_I}\,\, b_I\}.\,\,\,\,\,\,\,\, 
\end{eqnarray}
Note that the energy of the bosonic excitations in the normal phase can be obtained in the thermodynamical limit ($N_k\rightarrow \infty$) by neglecting the terms proportional to $(1/N_k)$ (for $k=C,I$) in the last Hamiltonian which therefore becomes  quadratic in the three bosonic modes $a, b_I$ and $b_C$. The frequency of the lower energy mode  can vanish for $\Omega_C=\Omega_C^{cr}=(1/2)\sqrt{\omega_{cav}\omega^0_C}$ and for $\Omega_I=\Omega_I^{cr}=(1/2)\sqrt{\omega_{cav}\omega^0_I}$, hence we have here two straight critical lines (one horizontal and one vertical) depicted in Fig. \ref{phasediagram}.

\begin{figure}
\begin{center}
\includegraphics[width=260pt]{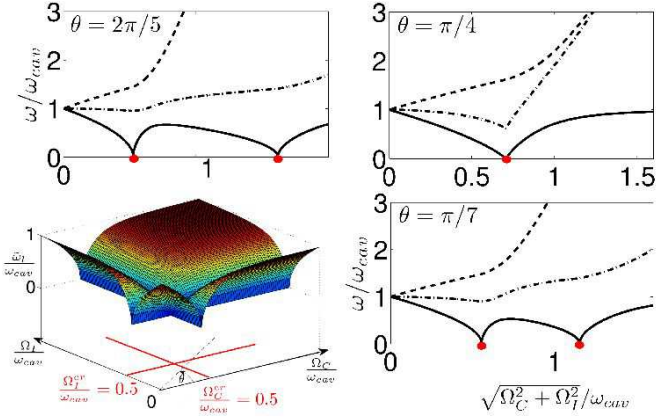}
\caption{ Frequencies of the $3$ branches of bosonic excitations (called lower, middle and upper) $\tilde{\omega}_l \leq  \tilde{\omega}_m  \leq  \tilde{\omega}_u$   for the resonant case $\omega^0_I=\omega^0_C=\omega_{cav}$ and  in units of $\omega_{cav}$.  3D surface plot: lower eigenmode $\tilde{\omega}_l$ as a function of $\Omega_C$ and $\Omega_I$. The other plots depict the three excitation frequencies versus $\sqrt{\Omega_C^2+\Omega_I^2}/\omega_{cav}$ for particular values of the polar angle in the parameter plane, namely $\theta =2\pi/5$, $\theta=\pi/4$ (up) and $\theta=\pi/7$ (down). For $\theta\not\equiv\pi/4\,\,\,[\pi/2]$, two different critical points mean two consecutive symmetry breakings. For $\theta\equiv\pi/4\,\,\,[\pi]$, the two symmetry breakings occur at the same time, implying a single quantum critical point. }
\label{excitations}
\end{center}
\end{figure}

To determine the phase diagram, we express the operators as
$b_k^{\dag} \rightarrow d_k^{\dag}\,-\,\sqrt{\beta_k}\,\,$($k=C,I$) and  $a^{\dag} \rightarrow c^{\dag}\,+\,\sqrt{\gamma_C}+i\sqrt{\gamma_I}$ 
 where  $\gamma_C$, $\gamma_I$, $\beta_C$ and $\beta_I$  are real  and such that  $\gamma_C \sim \beta_C \sim N_C $ and $\gamma_I \sim \beta_I \sim N_I $ \cite{emary}.
 By making the corresponding substitution in Eq.\ref{Hbosonic} and developing the square roots $(1 - (d_k^{\dag}\,-\,\sqrt{\beta_k})(d_k\,-\,\sqrt{\beta_k})/N_k)^{1/2}$,  we get an Hamiltonian with constant, linear and quadratic functions of the bosonic operators $c$, $d_C$, $d_I$
 and their hermitian conjugates. By setting the linear terms to zero \cite{emary} we find the solutions for the spontaneous mean fields: 
 \begin{eqnarray}
\sqrt{\gamma_C}+i\sqrt{\gamma_I}=\epsilon_C\frac{\Omega_C\sqrt{N_C(1-\tilde{\mu}_C^2)}}{\omega_{cav}}+i\epsilon_I\frac{\Omega_I\sqrt{N_I(1-\tilde{\mu}_I^2)}}{\omega_{cav}}\nonumber\\
\sqrt{\beta_C}=\epsilon_C\sqrt{\frac{N_C}{2}(1-\tilde{\mu}_C)}\,\,\text{;}\,\,\sqrt{\beta_I}=\epsilon_I\sqrt{\frac{N_I}{2}(1-\tilde{\mu}_I)}.\,\,\,\,
\end{eqnarray}
To write the results in a compact way, note that by definition  $\tilde{\mu}_k = 1$ if $\Omega_k<\Omega_k^{cr}$; $\tilde{\mu}_k = \frac{\omega^0_k\omega_{cav}}{4\Omega_k^2}$ if $\Omega_k>\Omega_k^{cr}$ , while $\epsilon_k=\pm1$.
These equations predict the existence of $4$ different phases. For $\Omega_C<\Omega_C^{cr}$ and  $\Omega_I<\Omega_I^{cr}$,
the {\it normal phase} is characterized by a non-degenerate vacuum with no macroscopic coherence ($\gamma_C = \gamma_I = \beta_C = \beta_I = 0$). For $\Omega_C>\Omega_C^{cr}$ and  $\Omega_I<\Omega_I^{cr}$, there is a {\it superradiant} phase (like in the Dicke model) with a twice degenerate vacuum with a real boson coherence $\langle a \rangle$ and a macroscopic spin polarization (along the $x$-direction) for the chain labeled by $C$, which is coupled to the quadrature $(a+a^{\dag})$. The third phase corresponds to
$\Omega_C<\Omega_C^{cr}$ and  $\Omega_I>\Omega_I^{cr}$: it is also {\it superradiant} with a twice degenerate vacuum with imaginary boson coherence and a macroscopic polarization of the chain of pseudo-spins labeled by $I$, which is coupled to the quadrature $i(a-a^{\dag})$.
Finally, when  $\Omega_C>\Omega_C^{cr}$ and  $\Omega_I>\Omega_I^{cr}$, the phase is {\it doubly superradiant} with a fundamental subspace that is $4$ times degenerate, with complex bosonic coherence and a macroscopic polarization for both the two independent spin chains.  Such a phase diagram is summarized in Fig. \ref{phasediagram}.  

In order to determine the energies of the bosonic excitations for the different phases, i.e. for any value of $\Omega_I$ and $\Omega_C$,  one has to diagonalize a  bosonic Hamiltonian quadratic in the fields  $c$, $d_C$, $d_I$ and their hermitian conjugates. There are $3$ branches of excitations whose positive frequencies ($\tilde{\omega}_l \leq  \tilde{\omega}_m  \leq  \tilde{\omega}_u$) are obtained by diagonalizing a $6\times6$ Bogoliubov matrix $\tilde{\mathcal{M}}=\begin{pmatrix} \tilde{\mathcal{P}}&\tilde{\mathcal{Q}}\\-\tilde{\mathcal{Q}}^{\dag}&-\tilde{\mathcal{P}}^T\end{pmatrix}$ where the $3\times3$ hermitian submatrix $\tilde{\mathcal{P}}$ and the $3\times3$ symmetric submatrix $\tilde{\mathcal{Q}}$ read:

\begin{eqnarray}
\tilde{{\mathcal{P}}}=\left (
  \begin{array}{ccc} \omega_{cav} & \tilde{\Omega}_C & i \tilde{\Omega}_I \\ \tilde{\Omega}_C & \tilde{\omega}^0_C+2\tilde{D}_C & 0 \\ -i \tilde{\Omega}_I & 0 & \tilde{\omega}^0_I + 2\tilde{D}_I\end{array}
  \right)
  \end{eqnarray} 
  \begin{eqnarray}
  \tilde{{\mathcal{Q}}}=\left (
  \begin{array}{ccc} 0 & -\tilde{\Omega}_C & -i \tilde{\Omega}_I\\-\tilde{\Omega}_C & -2\tilde{D}_C & 0 \\  -i \tilde{\Omega}_I & 0 & -2\tilde{D}_I  \end{array}
  \right),
\end{eqnarray} 
with $\tilde{\omega}^0_k=\omega^0_k(1+\tilde{\mu}_k)/(2\tilde{\mu}_k)$, $\tilde{\Omega}_k=\sqrt{2}\Omega_k\tilde{\mu}_k/\sqrt{1+\tilde{\mu_k}}$ and $\tilde{D}_k=\{\omega^0_k(3+\tilde{\mu}_k)(1-\tilde{\mu}_k)\}/(8\tilde{\mu}_k+8\tilde{\mu}_k^2)$ ($k=C,I$).
The values of the excitation frequencies as a function of $(\Omega_C,\Omega_I)$ are shown in Fig. \ref{excitations}. The behavior of $\tilde{\omega}_l$ at the quantum critical points is analogous to the one in the standard single-chain Dicke model \cite{emary}. The  double critical point $(\Omega_C^{cr},\Omega_I^{cr})$ behaves differently as the frequency vanishes linearly with the coupling.
The general expression (valid for any phase) for the ground state energy reads:
\begin{align}
\tilde{E}_G&=1/2(\tilde{\omega}_l+\tilde{\omega}_m+\tilde{\omega}_u-\omega_{cav}-\tilde{\omega}^0_C-\tilde{\omega}^0_I)\nonumber\\&-\sum_{k=C,I}\frac{\omega_k^0}{4\tilde{\mu}_k}\{N_k(1+\tilde{\mu}^2_k)+(1-\tilde{\mu}_k)\}.
\end{align}
 
It is also interesting to show a very simple asymptotic expression of the 4 degenerate vacua of the `double superradiant phase', in the finite size case, which we have derived in the limit $N_k\ll(\Omega_k/\omega_{cav})^2$ ($k=C,I$):
  \begin{eqnarray}
 &|G_{\pm+}\rangle& \simeq |\mp\frac{\Omega_C}{\omega_{cav}}\sqrt{N_C}+ i\frac{\Omega_I}{\omega_{cav}}\sqrt{N_I} \rangle\otimes |\pm\frac{N_C}{2},+\frac{N_I}{2}\rangle_x\,\,\,\,\nonumber \\
  &|G_{\pm-}\rangle& \simeq |\mp\frac{\Omega_C}{\omega_{cav}}\sqrt{N_C}- i\frac{\Omega_I}{\omega_{cav}}\sqrt{N_I} \rangle\otimes |\pm\frac{N_C}{2},-\frac{N_I}{2}\rangle_x\,\,\,\,.
 \end{eqnarray}  
which is a product of a coherent state for the boson part times two {\it ferromagnetic}  states polarized in the pseudospin $x$-direction for the two chains, namely $|\pm N_C/2\rangle_x\otimes|\pm N_I/2\rangle_x$. Those states satisfy $J_x^k|\pm N_k/2\rangle_x=\pm(N_k/2)|\pm N_k/2\rangle_x$ ($k=C,I$),  meaning that each atom is maximally polarized. 
\begin{figure}
\begin{center}
\includegraphics[width=290pt]{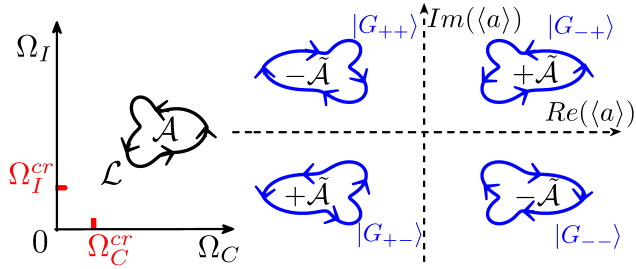}
\caption{Left: Closed loop $\mathcal{L}$ enclosing an area $\mathcal{A}$ in the 2D parameters plane $(\Omega_C,\Omega_I)$.
If the trajectory occurs in the part of the phase diagram with the four-fold degenerate ground state, then the corresponding adiabatic transformation produces a non-abelian Berry unitary transformation in the basis $\{|G_{++}\rangle,|G_{+-}\rangle,|G_{-+}\rangle,|G_{--}\rangle\}$.}
\label{berry}
 \end{center}
 \end{figure}
For a finite number of spins,  the exact degeneracy is lifted,  the first 4 eigenstates being linear superpositions of the $4$ states $|G_{\pm\pm}\rangle$. The 3 energy splittings which appear  exponentially decrease with the size and/or the coupling constants, being either $\sim \exp(-2N_C \Omega_C^2/\omega_{cav}^2)$  or $\sim \exp(-2N_I \Omega_I^2/\omega_{cav}^2)$ in analogy to the standard Dicke model\cite{degvacua}, implying that for many applications they can be neglected. 

An interesting property of the present system is the possibility of manipulating the four-fold ground space via geometric Berry effects.
If we vary $\Omega_C$ and $\Omega_I$ to form a loop trajectory $\mathcal{L}$ in the corresponding two-dimensional parameter space (as shown in Fig. \ref{berry}) in an adiabatic fashion (slowly with respect to $1/\omega_{cav}$), then one creates an non-abelian Berry unitary transformation \cite{nonabelian} $U(A)$ in the basis $\{|G_{++}\rangle,|G_{+-}\rangle,|G_{-+}\rangle,|G_{--}\rangle\}$. Namely:
\begin{eqnarray}
U(\tilde{\mathcal{A}})=\mathcal{P}e^{-\oint_{\mathcal{L}}\sum_{k=I,C}\langle G_{\pm\pm}|\frac{\partial}{\partial \Omega_k} |G_{\pm\pm}\rangle d\Omega_k}\simeq e^{i\tilde{\mathcal{A}}\Sigma_{z}^C\otimes \Sigma_{z}^I}\,\,\,\,\,\,\,\,\,\,
\end{eqnarray}
 where the geometric angle is $\tilde{\mathcal{A}}\simeq 2\mathcal{A} \sqrt{N_C N_I}/\omega_{cav}^2$ with $\mathcal{A}$ the loop area in the parameter plane $(\Omega_C,\Omega_I)$ (see Fig. \ref{berry}) and $\Sigma_{z}^C\otimes\Sigma_{z}^I=diag(1,-1,-1,1)$. The path ordering operator $\mathcal{P}$ had been be taken away because the Wilczek-Zee connections\cite{nonabelian} $\langle G_{\pm\pm}|\frac{\partial}{\partial \Omega_k} |G_{\pm\pm}\rangle$ are almost diagonal since  $\exp(-2N_k \Omega_k^2/\omega_{cav}^2)\ll 1$ (for $k=I,C$).  Note that if one sees the two-chain system as a two-qubit system, then such non-abelian transformation is equivalent to a conditional two-qubits quantum gate.  
We would like to point out that we have successfully tested our analytical results with finite-size exact diagonalizations.  
 \begin{figure}
\begin{center}
\includegraphics[width=220pt]{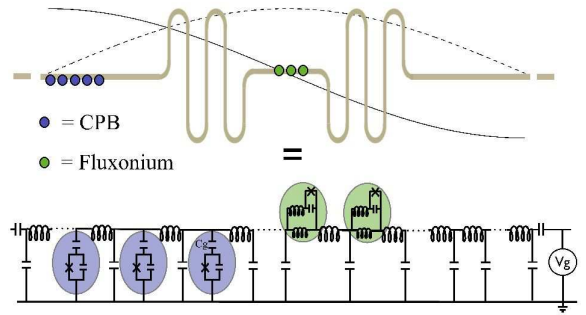}
\caption{A possible realization of the Hamiltonian \ref{hamiltonian} in circuit QED. One chain of Cooper pair boxes (blue) capacitively coupled to the voltage bosonic field of a transmission line resonator and one chain of fluxoniums (green) inductively coupled to the flux bosonic field of the resonator. Because of the boundary conditions\cite{Wallraff}, the first mode profile of the voltage (resp. current) field has its antinodes at the edges (resp. center) of the resonator, as shown above in solid (resp. dashed) line. The capacitive coupling provides a coupling to a quadrature of the resonator bosonic field, while the inductive coupling provides the coupling to the other quadrature.}
\label{sketch}
\end{center}
\end{figure}

The double chain Dicke model may be realized in circuit QED systems, by considering, for example, a chain of Josephson two-level atoms capacitively coupled to a transmission line resonator and considering another chain of Josephson atoms inductively coupled to the same resonator. The `capacitive' and `inductive'  schemes are two different ways of coupling a two-level system in a transmission line resonator in circuit QED\cite{devoretstrong}. The capacitive coupling connects the charges of a Josephson two-level system to the resonator voltage quantum field $V_r\propto (a +a^{\dag})$ \cite{Wallraff}. The inductive coupling, which is dual of the latter , provides an interaction between the flux of the Josephson atom and the current of the resonator  which is proportional to the other quadrature of the field, i.e. $I_r\propto i(a-a^{\dag})$, because the resonator current is proportional to the time derivative of the resonator voltage.
 In particular, a chain of $N_I$ identical fluxonium Josephson atoms\cite{devoret}, inductively coupled to the first bosonic mode of the resonator can undergo a spin-boson interaction of the form $H_I/\hbar = 2i\Omega_I/\sqrt{N_I}(a -a^{\dag})J_x^I$ \cite{degvacua}.  The coupling Hamiltonian of a collection of $N_C$ identical Cooper pair boxes\cite{nakamura},  interacting with the first bosonic mode of the resonator, with a gate capacitance $C_g$ and a static gate voltage $V_g$ such that $n_g=C_gV_g/(2e)=1/2$ can be written as  $H_C/\hbar = 2\Omega_C/\sqrt{N_C}(a + a^{\dag})J_x^C$ \cite{cpb,Acarre}.
One way to combine these two kinds of couplings and get the Hamiltonian \ref{hamiltonian} is sketched in Fig. \ref{sketch}. 
Note that in such a context, the transformation $\mathcal{T}_I$ might be defined as the time reversal symmetry since it lets unchanged the resonator voltage and the qubit charge operator  while it flips the resonator current  and the qubit flux. And  $\mathcal{T}_C$ will be then the product of the parity transformation $\Pi$ and the  time reversal symmetry $\mathcal{T}_I$.
Finally, this kind of Hamiltonians could also be implemented with atomic Bose-Einstein condensates dressed by judiciously tailored laser pump beams in an optical cavity \cite{ritsch} in a configuration related to the recent observation of the dynamic Dicke quantum phase transition\cite{ZurichNature,ZurichPRL}.
  
In conclusion, a spin-boson system with a quantum phase transition characterized by two independent symmetry breakings has been presented. The full phase diagram of such `double chain' Dicke model has been solved, showing the possibility to obtain four-fold vacuum degeneracy associated to  a `doubly superradiant phase'. We have shown how to manipulate the ground states via non-abelian Berry methods. Our work provides a new paradigm in the very active field of spin-boson systems, with possible applications in circuit QED and cavity QED with superfluid atoms.
We wish to thank K. Le Hur, J. P. Gazeau, H. Ritsch, D. Hagenm\"uller for fruitful discussions. C. C. is member of {\it Institut Universitaire de France}.


\begin{thebibliography}{}
\bibitem{Acarre} The capacitive coupling also leads to the appearance of a quadratic field of the form $\bar{D}(a\,+\,a^{\dag})^2$ , which can be rendered arbitrarily small under conditions for example given in \cite{cpb}.
\end{thebibliography}
  \end{document}